\documentclass[showpacs,amsmath,amssymb]{revtex4}
\usepackage{graphicx}
\usepackage{dcolumn}
\usepackage{bm}
\usepackage{epsfig}

\begin{document}
\newcommand{\etal}{\it et al. \rm}
\newcommand{\pp}{\psi(2S)}
\newcommand{\jp}{J/\psi}
\newcommand{\pppp}{\psi(2S) \rt \pi^+ \pi^- J/\psi}
\newcommand{\ppll}{\psi^{'} \rightarrow
 \pi^+ \pi^- J/\psi$, $J/\psi \rightarrow \ell^+ \ell^-}
\newcommand{\ppmm}{\psi^{'} \rightarrow
 \pi^+ \pi^- J/\psi$, $J/\psi \rightarrow \mu^+ \mu^-}
\newcommand{\ppee}{\psi^{'} \rightarrow
 \pi^+ \pi^- J/\psi$, $J/\psi \rightarrow e^+ e^-}
\newcommand{\nppj}{N_{\pi^+ \pi^- J/\psi}}
\newcommand{\eppj}{\epsilon_{\pi^+ \pi^- J/\psi}}
\newcommand{\rt}{\rightarrow}
\newcommand{\bi}{\begin{itemize}}
\newcommand{\ei}{\end{itemize}}
\newcommand{\im}{\item}
\newcommand{\be}{\begin{enumerate}}
\newcommand{\ee}{\end{enumerate}}

\preprint{Draft-PRD}

\parindent = 0.5 in


\title{\boldmath Study of $\pp$ decays to $X J/\psi$ }

\author{
M.~Ablikim$^{1}$, J.~Z.~Bai$^{1}$, Y.~Ban$^{10}$, 
J.~G.~Bian$^{1}$, X.~Cai$^{1}$, J.~F.~Chang$^{1}$, 
H.~F.~Chen$^{16}$, H.~S.~Chen$^{1}$, H.~X.~Chen$^{1}$, 
J.~C.~Chen$^{1}$
% ...
\\(BES Collaboration)\\ 
\vspace{0.2cm}
$^1$ Institute of High Energy Physics, Beijing 100039, People's Republic of China\\
% ...
$^{10}$ Peking University, Beijing 100871, People's Republic of China\\
% ...
$^{16}$ University of Science and Technology of China, Hefei 230026, People's Republic of China\\
% ...
}

\date{\today}
\begin{abstract}
Using $J/\psi \rt \mu^+ \mu^-$ decays from a sample of approximately
$4 \times 10^6$ $\pp$ events collected with the BESI detector, the
branching fractions of $\pp \rt \eta J/\psi$, $\pi^0 \pi^0 J/\psi$,
and anything $J/\psi$ normalized to that of $\pp \rt \pi^+ \pi^-
J/\psi$ are measured.  The results are $B(\pp \rt \eta J/\psi)/B(\pp
\rt \pi^+ \pi^- J/\psi) = 0.098 \pm 0.005 \pm 0.010$, $B(\pp \rt
\pi^0 \pi^0 J/\psi)/B(\pp \rt \pi^+ \pi^- J/\psi) = 0.570 \pm 0.009
\pm 0.026$, and $B(\pp \rt {\rm anything} \: J/\psi)/B(\pp \rt \pi^+
\pi^- J/\psi) = 1.867 \pm 0.026 \pm 0.055$.
\end{abstract}


\maketitle


\section{Introduction}

Transitions of the type $\pp \rt X J/\psi$ comprise a large fraction
of the total $\pp$ decay width.  They include exclusive decays where
$X = \eta$, $\pi^0,$ and $\pi \pi$, as well as the cascade processes
$\pp \rt \gamma \chi_{c0/1/2}$, $\chi_{c0/1/2} \rt \gamma J/\psi$.
The inclusive branching fraction is measured to be $B(\pp \rt {\rm
  anything} \: J/\psi) = (55.7 \pm 2.6) \%$; the contributions from
the individual sub-processes are less precisely known~\cite{PDG02}.

These branching fractions are important in
understanding the hadronic decay dynamics of vector
charmonia~\cite{suzuki,gu}, since the inclusive hadronic decay
branching fraction is calculated by subtracting them
from unity. Present branching fractions may indicate
a possible excess of the $\psi(2S)$ hadronic decay rate relative to 
the "12\% rule" prediction from $J/\psi$ decays.

Recently the BES experiment reported new measurements with
improved precision for $B(\pp \rt \pi^0 J/\psi)$, $B(\pp \rt \eta
J/\psi)$, and $B(\pp \rt \gamma \chi_{c0/1/2})B(\chi_{c0/1/2}\rt
\gamma J/\psi)$ using decays $\pp \rt \gamma \gamma J/\psi, J/\psi \rt
\ell^+ \ell^-$, (where $\ell^+\ell^- =\mu^+\mu^-$ or
$e^+e^-$)~\cite{ggjp}.  Previous measurements are few and date back to
the 1970's and 80's~\cite{markI, CNTR, DASP, markii, crystalball}.
More precise measurements are needed.

In this paper, we report the results of a different technique for
measuring branching fractions for the inclusive decay $\pp \rt {\rm
  anything} \: J/\psi$, and the exclusive processes for the cases
where $X = \eta$ and $X = \pi\pi$.  We reconstruct $\mu^+ \mu^-$ pairs
and determine the number of $\pp \rt X J/\psi$ events in our data
sample from the $J/\psi \rt \mu^+ \mu^-$ peak in the $\mu^+ \mu^-$
invariant mass distribution (see Fig.~\ref{fig:fit_mmumu}).  The
exclusive branching fractions are determined from fits to the
distribution of masses recoiling from the $J/\psi$ with Monte-Carlo
determined distributions for each individual channel. We distinguish
the $\pi^+ \pi^-$ and $\pi^0 \pi^0$ contributions from simultaneous
fits to event samples with and without accompanying charged particles.
To avoid a number of systematic errors, the channels of interest are
normalized to the observed number of $\pi^+ \pi^- J/\psi$ events; we
report ratios of the studied branching fractions to that for $B(\pp
\rt \pi^+ \pi^- J/\psi)$.  This analysis is based on a sample of
approximately $4 \times 10^6$ $\pp$ events obtained with the Beijing
Spectrometer detector (BESI)~\cite{besI} at the Beijing
Electron-Positron Collider (BEPC).

\section{BESI detector and Monte Carlo Simulations}

The Beijing Spectrometer, BESI, is a conventional cylindrical magnetic
detector that is coaxial with the BEPC colliding $e^+e^-$ beams.  It
is described in detail in Ref.~\cite{besI}. A four-layer central drift
chamber (CDC) surrounding the beam pipe provides trigger information.
Radially outside of the CDC, a forty-layer main drift chamber (MDC)
provides tracking and energy-loss ($dE/dx$) information on charged
tracks over $85\%$ of the total solid angle.  The momentum resolution
is $\sigma _p/p = 1.7 \% \sqrt{1+p^2}$ ($p$ in GeV/c), and the $dE/dx$
resolution for hadron tracks for this data sample is $\sim 9\%$.  An
array of 48 scintillation counters surrounding the MDC provides
measurements of the time-of-flight (TOF) of charged tracks with a
resolution of $\sim 450$ ps for hadrons.  Outside of the TOF system is
a 12 radiation length lead-gas barrel shower counter (BSC), operating
in self-quenching streamer mode, that measures the positions and
energies of electrons and photons over 80\% of the total solid angle.
The energy resolution is $\sigma_E/E= 22 \%/\sqrt{E}$ ($E$ in GeV).
Surrounding the BSC is a solenoid magnet that provides a 0.4 Tesla
magnetic field in the central tracking region of the detector. Three
double layers of proportional chambers instrument the magnet flux
return (MUID) and are used to identify muons with momentum greater
than 0.5 GeV/c.


Monte Carlo simulations are used to determine efficiencies and the expected
recoil mass distributions for the various processes involved, including 
$\psi(2S) \rt \gamma \chi_{c1/2}$, $\psi(2S) \rt \pi \pi J/\psi$, and
$\psi(2S) \rt \eta J/\psi$ with $J/\psi \rt \mu^+ \mu^-$, as well as
the background processes $e^+ e^- \rt \gamma \mu^+ \mu^-$, $e^+ e^-
\rt \pp$, $\pp \rt (\gamma) \mu^+ \mu^-$, $e^+ e^- \rt 2 \gamma^* \rt
\mu^+ \mu^- e^+ e^-$ $(\mu^+ \mu^- \mu^+ \mu^-)$. In each decay,
angular distributions are generated according to expectations for
that process.  The agreement between the distributions of $\cos
\theta_{\mu}$ for data and Monte Carlo has been checked in separate
analyses and found to be reasonable.  Since the default generator for
$\psi(2S) \rt \pi \pi J/\psi$, produces $S$-wave dipion states,
while BES has measured a small but non-negligible amount of $D$-wave in 
the dipion system
\cite{distributions}, Monte Carlo events are weighted to give the
correct angular and $m_X$ distributions.

\section{Event selection}

Selected events are required to have 
more than one and less than six charged tracks.

\subsection{Muon Selection}

Events must have two identified 
muon tracks with zero net charge.  The $\mu$ tracks
must satisfy:

\begin{enumerate}

\im $0.5 < p_{\mu} < 2.5$ GeV/c.  Here $p_{\mu}$ is the three-momentum
of the candidate muon track in the lab.

\im $p_{\mu^+} > 1.3$ or $p_{\mu^-}> 1.3$ GeV/c or $(p_{\mu^+} +
p_{\mu^-}) > 2.4$ GeV/c.  This requirement selects events consistent with
$J/\psi$ decay, while rejecting background.

\item $|\cos\theta_{\mu}|<0.60$.  Here $\theta_{\mu}$ is the
laboratory polar angle of the muon.  This requirement ensures that muons are
contained in the MUID system.

\item $\cos \theta_{\mu^+ \mu^-} < -0.85$. This is the cosine of
the angle between the two leptons in the lab.  The leptons from this
decay are almost back to back.
 

\item Both tracks must have $N^{hit} > 1$, where $N^{hit}$ is the
number of MUID layers with matched hits and ranges from 0 to 3.

\im $|t_{\rm TOF}(\mu^+) - t_{\rm TOF}(\mu^-)| < 4$ ns. Here $t_{\rm
  TOF}$ is the time measured by the TOF counters.  This requirement
  removes cosmic ray background.

\end{enumerate}

\noindent


\subsection{\boldmath Selection of $J/\psi$ events}

\begin{figure}[!htb]
\centerline{\epsfysize 5.0 truein
\epsfbox{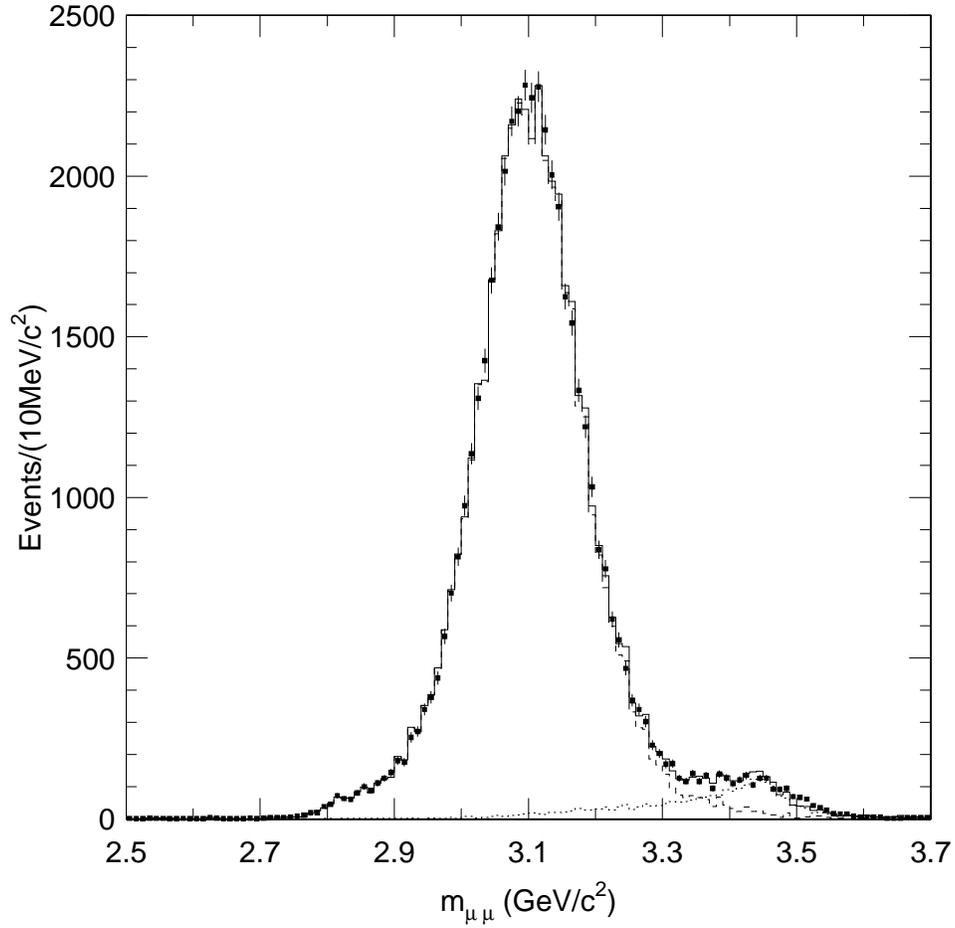}}
\caption{\label{fig:fit_mmumu} Distribution of dimuon invariant mass,
  $m_{\mu \mu}$, for events that pass the $J/\psi \rt  \mu^+ \mu^-$  
  kinematic fit. Dots with error bars are data.  Also shown is the fit
  (solid histogram) to the distribution with signal (long dashed
  histogram) and background (short dashed histogram) shapes.  }
\end{figure}

\begin{figure}[!htb]
\centerline{\epsfysize 2.5 truein
\epsfbox{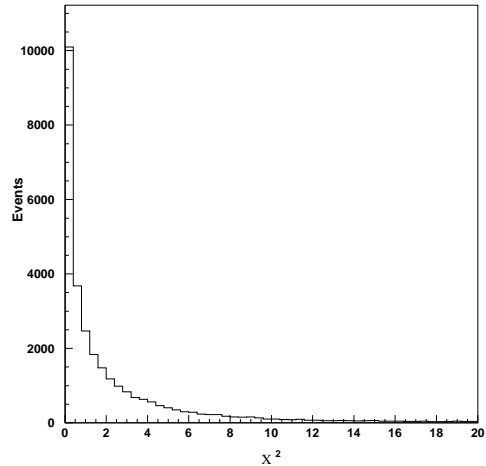}}
\caption{\label{fig:chisquare}
Distribution of $\chi^2$ (data) for events satisfying 
a one-constraint kinematic fit to $J/\psi \rt
\mu^+ \mu^-$.}
\end{figure}

For $J/\psi \rt \mu^+ \mu^-$ candidates, the two tracks must satisfy a one
constraint kinematic fit to the $J/\psi$ mass.  Shown in
Fig.~\ref{fig:fit_mmumu} is the invariant mass distribution of the two
muons, $m_{\mu \mu}$, for $J/\psi$ candidates.  A clear peak at the
$J/\psi$ mass is evident above background. 
The distribution of $\chi^2$ values from the one-constraint
kinematic fits to the $J/\psi\rt\mu^+\mu^-$
hypothesis is shown in Fig.~\ref{fig:chisquare}.
The mass recoiling against the $J/\psi$ candidates, $m_X$ is determined
from energy and momentum conservation.

\subsection{\boldmath Extra track ($\pi$) selection}
In order to distinguish $\psi(2S) \rt \pi^+ \pi^- J/\psi$ and
$\psi(2S) \rt \pi^0 \pi^0 J/\psi$ events, separate $m_X$ histograms
are made for events with no additional charged tracks, 
those with any number of additional charged
tracks, and those with two or more additional charged tracks.
The first histogram and one of the other histograms  
are fitted simultaneously \cite{mnfit}. 
To reduce background and improve the quality of the track momentum
measurements,
events used for this part of the analysis are required to have a
kinematic fit $\chi^2 < 7$.

Additional charged tracks can originate from
$\psi(2S) \rt \pi^+ \pi^- J/\psi$ and $\psi(2S) \rt \eta J/\psi$,
$\eta \rt \pi^+ \pi^- \gamma/\pi^0$ decays, and also from
gamma conversions and delta rays.
Selection criteria are applied to the additional tracks  
to enhance the selection of low energy pion tracks, such as 
those coming from the process $\psi(2S) \rt \pi^+ \pi^- J/\psi$,
and reject gamma conversions and delta rays.
\be
\im
Tracks must have a good helix fit with:
\item $p_{\pi} < 0.5$ GeV/c,
where $p_{\pi}$ is the pion momentum;
\item $|\cos \theta_{\pi}| < 0.8$,
where $\theta_{\pi}$ is the polar angle of the $\pi$ in the laboratory 
system;
\item $p_{xy_{\pi}} > 0.08$ GeV/c,
where $p_{xy_{\pi}}$ is the momentum of the pion transverse to the beam
direction (this removes tracks that circle in the MDC;
\item $|\chi^{dE/dx}_{\pi}| < 3.0$. $\chi^{dE/dx}_{\pi} = \frac{(dE/dx)_{meas}
 - (dE/dx)_{exp}}{\sigma}$,
where $(dE/dx)_{meas}$ and $(dE/dx)_{exp}$ are the measured
and expected $dE/dx$ energy losses 
for pions, respectively, and $\sigma$ is the experimental $dE/dx$ resolution.
\item For events with  more than one additional tracks, 
$\cos\theta_{\pi \pi} < 0.9, $
where $\theta_{\pi \pi}$ is the laboratory angle between them.
This last requirement reduces contamination from misidentified
$e^+e^-$ pairs from $\gamma$ conversions.
\ee

\begin{figure}[!htb]
\centerline{\epsfysize 5.5 truein
\epsfbox{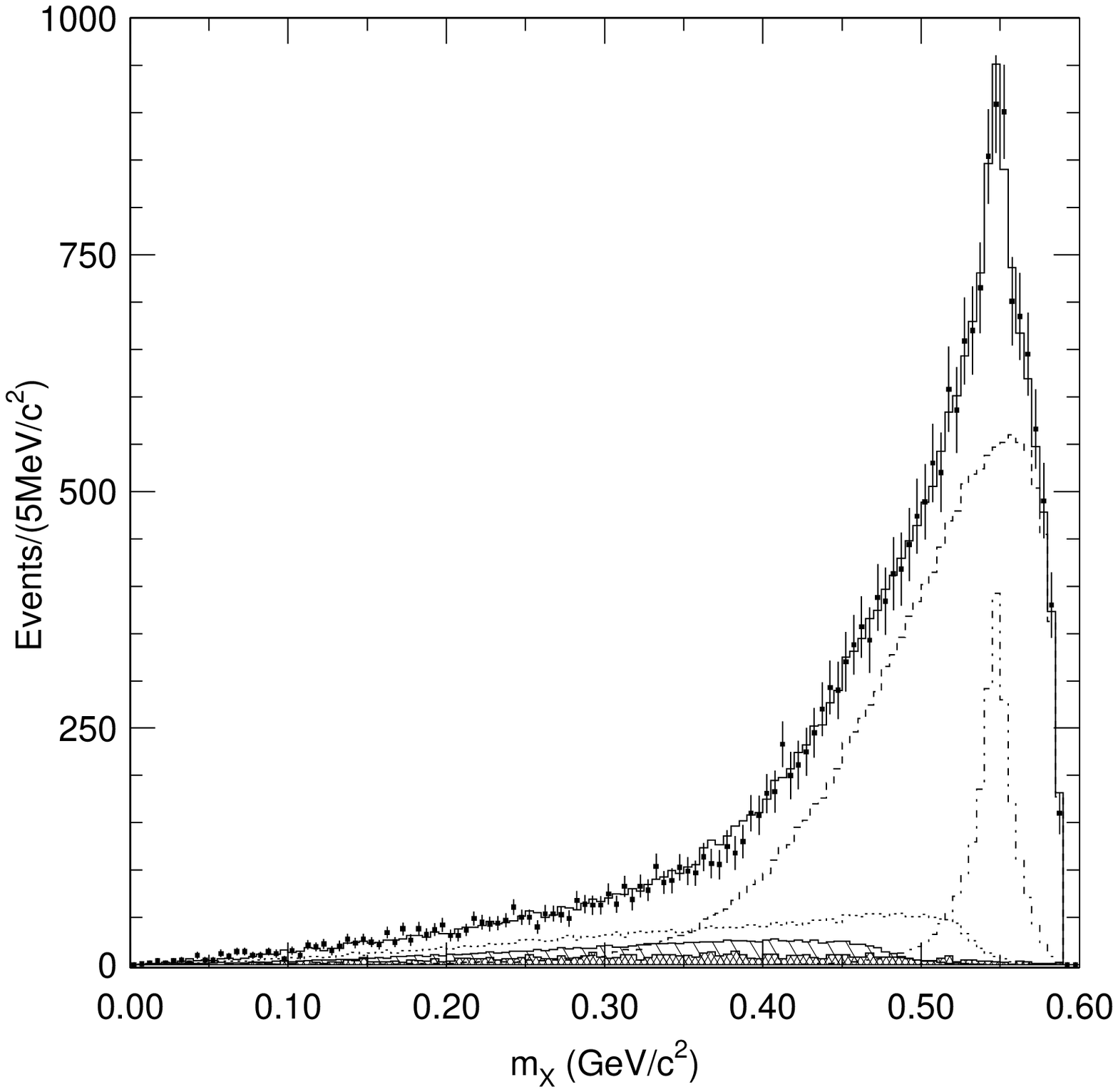}}
\caption{\label{fig:none} Fit of the $m_X$ distribution events with
no additional charged tracks.  Shown are the data (points with error 
bars), the component histograms, and the final fit.  For the components, 
the large, long-dash histogram is $\psi(2S) \rt \pi \pi J/\psi$, the
narrow, dash-dot histogram is $\psi(2S) \rt \eta J/\psi$, the broad,
short-dashed histogram is $\pp \rt \gamma \chi_{c1}$, the broad,
hatched histogram is $\pp \rt \gamma \chi_{c2}$, and the lowest
cross-hatched histogram is the combined $e^+ e^- \rt \gamma \mu^+
\mu^-$ and $e^+ e^- \rt \psi(2S), \psi(2S) \rt (\gamma)\mu^+ \mu^-$
background. The final fit is the solid histogram. }
\end{figure}

\begin{figure}[!htb]
\centerline{\epsfysize 5.5 truein
\epsfbox{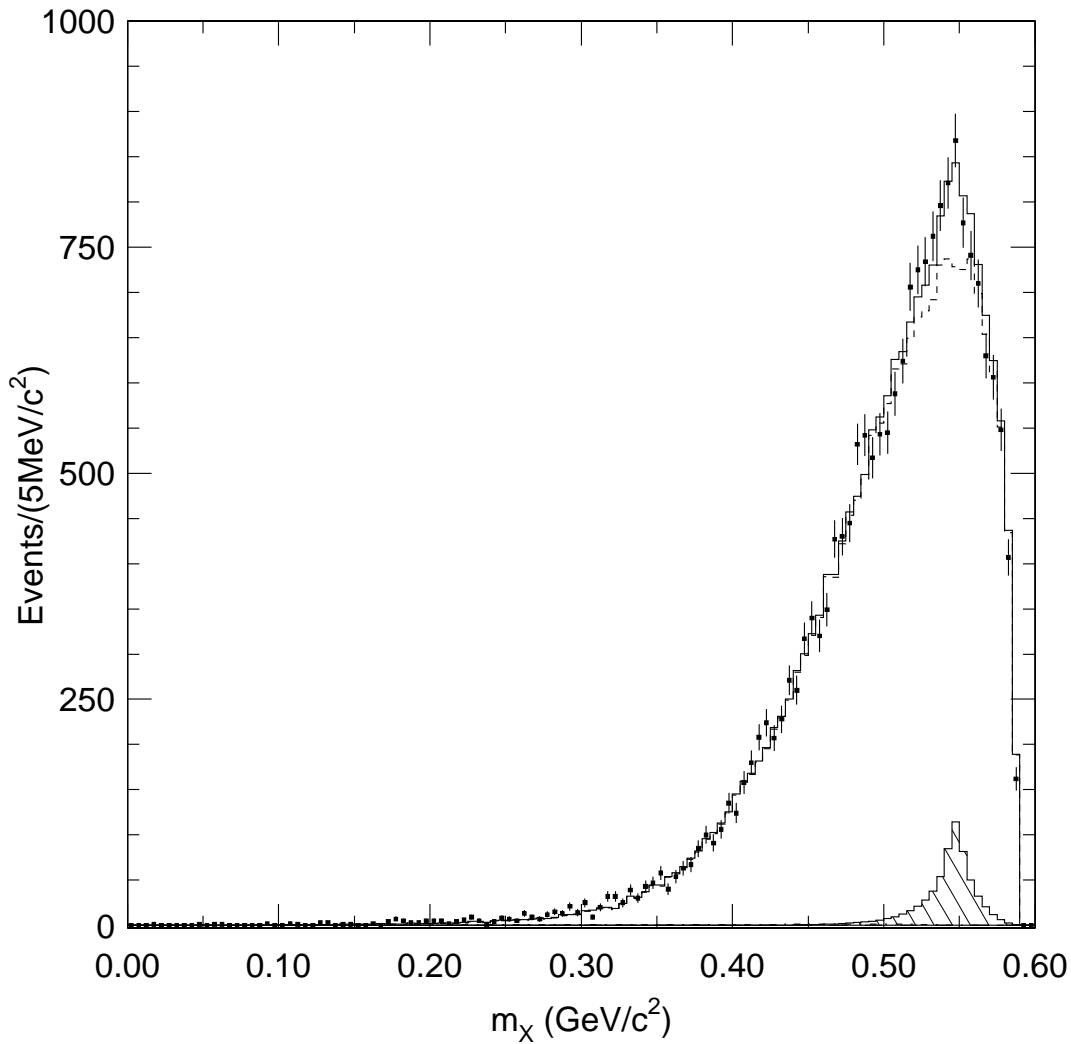}}
\caption{\label{fig:one} Fit of the $m_X$ distribution for events
with any number of additional charged tracks.  Shown are the data (points 
with error
bars), the component histograms, and the final fit (solid
histogram). The dashed histogram is  $\psi(2S) \rt \pi^+ \pi^-
J/\psi$, and the hatched histogram is  $\psi(2S) \rt \eta J/\psi$. 
There is very little evidence for $\chi_{c1/2}$.  This distribution is
composed predominantly of $\psi(2S) \rt \pi^+ \pi^- J/\psi$.  }
\end{figure}

The
$m_X$ histograms for events with and without additional charged
tracks, selected according to the above requirements, are shown in Figs.~\ref{fig:none}
and \ref{fig:one}.

\section{Background studies}

The backgrounds from $e^+ e^- \rt \gamma \mu^+ \mu^-$
and $e^+ e^- \rt \pp, \pp \rt (\gamma) \mu^+ \mu^-$ in the $m_{\mu
  \mu}$ and  $m_X$ distributions are measured using data.  We begin
by considering what backgrounds are expected.

\subsection{Expected backgrounds}

Possible background processes are $e^+ e^- \rt \gamma \mu^+ \mu^-$, $e^+
e^- \rt \pp, \pp \rt (\gamma) \mu^+ \mu^-$, and $e^+ e^- \rt 2 \gamma^*
\rt \mu^+ \mu^- e^+ e^- (\mu^+ \mu^- \mu^+ \mu^-)$.  The number of
background events expected from the simulation of $e^+ e^- \rt 2
\gamma^* \rt \mu^+ \mu^- e^+ e^-$ in the $m_{\mu
\mu}$ distribution is four events, which is negligible.  The
background in the $m_X$ distribution with no extra charged tracks is
even smaller, and the calculated backgrounds for $e^+ e^- \rt \mu^+
\mu^- \mu^+ \mu^-$ are determined to be smaller still.  The numbers of
two photon background events are found to be entirely negligible, and
this process is ignored further.

The Monte Carlo (MC)-determined level of $e^+ e^- \rt \gamma \mu^+ \mu^-$
background in the $m_{\mu \mu}$ \cite{mumu} distribution and the $m_X$ 
distribution with no additional tracks \cite{mx} are 556 $\pm$ 20,
and 192 $\pm$ 10 events, respectively.
For $e^+ e^- \rt \pp, \pp \rt (\gamma) \mu^+ \mu^-$, the expected background
in these distributions are 430 $\pm$ 67 and 137 $\pm$ 22 events,
and the total
backgrounds are 986 $\pm$ 70 and 329 $\pm$ 24 events, respectively.  
The result obtained from fitting the background level in the $m_{\mu \mu}$ 
distribution (Fig.~\ref{fig:fit_mmumu}) is 1307 $\pm$ 56 events
which is larger than the MC-determined background level. 
The background shapes and the numbers of background events estimated for
the two processes are similar, so the histograms are combined in the
fitting of the two distributions.

\subsection{Inclusive Background}

\subsubsection{Determination of $e^+ e^- \rt \gamma
  \mu^+ \mu^-$ background from the $\cos \theta_{J/\psi}$ distribution.}

\begin{figure}[!htb]
\centerline{\epsfysize 3.0 truein
\epsfbox{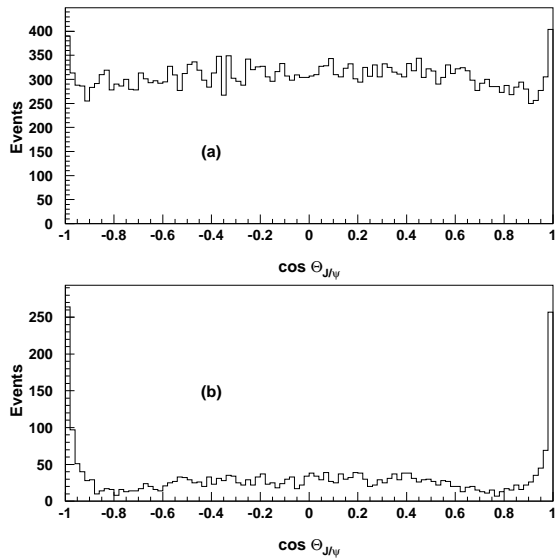}}
\caption{\label{fig:cospsi} The distribution of the cosine of the
angle of the $J/\psi$ in the lab, $\cos \theta_{J/\psi}$, for a.) data
and b.)  simulated $e^+ e^- \rt \gamma \mu^+ \mu^-$ events.  The peaks
at $|\cos \theta_{J/\psi}| = 1$ indicate some $e^+ e^- \rt \gamma
\mu^+ \mu^-$ background in the data.  No $\chi^2$ requirement is made for
these plots but $m_{\mu \mu} < 3.4$ GeV/c$^2$.  }
\end{figure}

The photon from $e^+ e^- \rt \gamma
\mu^+ \mu^-$ is typically emitted
along the beam direction, producing a dimuon system which is along the
beam in the opposite direction.  The cosine of the angle of
the $J/\psi$ in the lab, $\cos \theta_{J/\psi}$, shows a strong
peaking near $\pm 1$ for simulated $e^+ e^- \rt \gamma \mu^+ \mu^-$
events, as shown in Fig.~\ref{fig:cospsi}b.  Some peaking near $\pm 1$
is also found in the data, as shown for the case with no $\chi^2$
requirement in Fig.~\ref{fig:cospsi}a.  By fitting the $\cos
\theta_{J/\psi}$ distribution for data with the $e^+ e^- \rt \gamma
\mu^+ \mu^-$ Monte Carlo distribution plus a distribution to represent
the non-peaked events, a total of 839 $\pm$ 98 background events is
obtained.  This can be compared with the predicted $e^+ e^- \rt \gamma
\mu^+ \mu^-$ background estimate of 556 $\pm$ 20 events in the $m_{\mu \mu}$
distribution.  The difference is 283 $\pm$ 100. 

An additional source of background, not included in the
$e^+ e^- \rt \gamma \mu^+ \mu^-$ simulation, is due to ``radiative
return'' to the $J/\psi$ peak, $e^+ e^- \rt \gamma J/\psi,$ $J/\psi
\rt \mu^+ \mu^-$.  This background is similar to the signal in
the $m_{\mu^+ \mu^-}$ distribution and would not be part of the
background determined from the fit to the $m_{\mu^+ \mu^-}$
distribution.  However, it would be included in the background
determined from the $\cos \theta_{J/\psi}$ distribution.
The difference between the background determined from the $\cos
\theta_{J/\psi}$ distribution and the predicted $e^+ e^- \rt \gamma
\mu^+ \mu^-$ is taken as a measure of the radiative return.

\subsubsection{Total background from the  $\Delta
\phi$ distribution}

\begin{figure}[!htb]
\centerline{\epsfysize 4.5 truein
\epsfbox{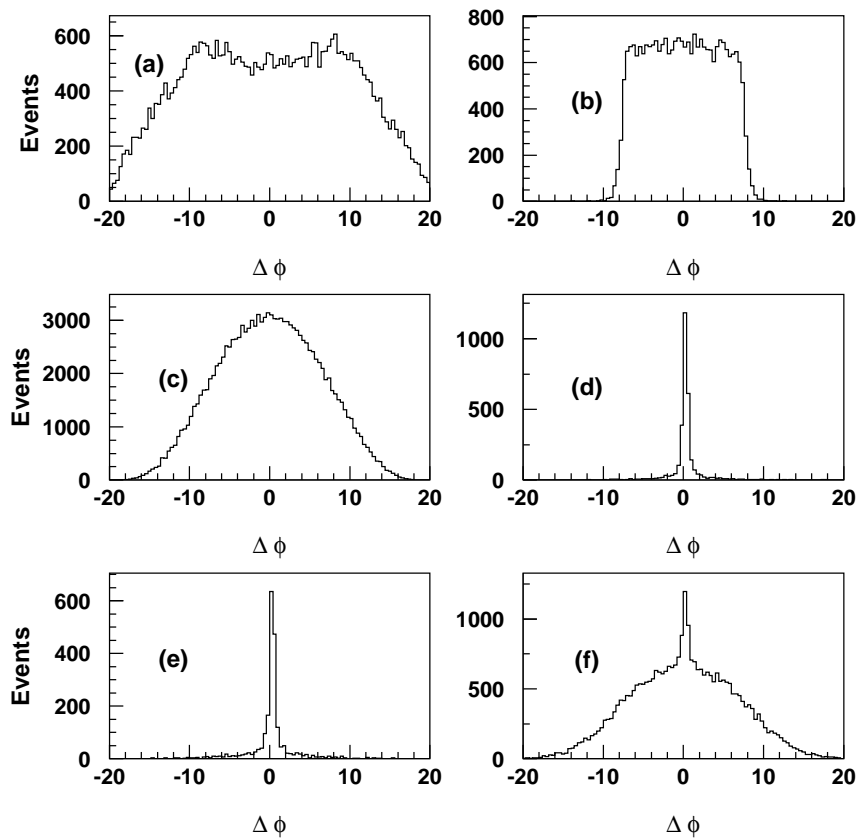}}
\caption{\label{fig:dphi} $\Delta \phi$ distribution for a.)  $\gamma
\chi_{c1}$, b.) $\eta J/\psi$, c.)  $\pi^+ \pi^- J/\psi$, d.) $e^+ e^-
\rt \gamma \mu^+ \mu^-$, e.)  $e^+ e^- \rt \psi(2S), \psi(2S) \rt
\gamma \mu^+ \mu^-$, and f.)  data. No $\chi^2$ requirement is made for these
plots.  }
\end{figure}

Since the $ \mu^+$ and $\mu^-$ from 
$e^+ e^- \rt \gamma \mu^+ \mu^-$, $e^+ e^- \rt \pp, \pp \rt (\gamma) \mu^+
 \mu^-$, and the ``radiative return'' background processes are 
coplanar, it is possible to determine their level from
the $\Delta\phi$ distributions of the data, where 
$\Delta\phi$ is the difference
 between the $\phi$ angles of the $\mu^+$ and the $\mu^- (\pm 180^o$).
Fig.~\ref{fig:dphi} shows the Monte Carlo
 and data distributions for $\Delta \phi$. 
 The $e^+ e^- \rt \gamma \mu^+ \mu^-$ and $e^+ e^- \rt \pp, \pp \rt
 (\gamma) \mu^+ \mu^-$ distributions show a large peak at $0^o$, which
 is not seen in other simulations.  A similar peak is seen in
 the data.  The $\Delta \phi$ distribution from data is fitted using
 Monte Carlo distributions, including the combined $e^+ e^- \rt \gamma
 \mu^+ \mu^-$ and $e^+ e^- \rt \pp, \pp \rt (\gamma) \mu^+ \mu^-$
 $\Delta \phi$ distribution plus a broad distribution to represent the
 other processes.  The fit determines $1476 \pm 150$ background events
 in the $m_{\mu \mu}$ distribution.  The result is summarized in
 Table~\ref{tab:final_iv}, along with estimates obtained by adding the
 ``radiative return'' determined using the $\cos \theta_{J/\psi}$
 distribution to the predicted and measured background from the
 $m_{\mu \mu}$ distribution.  The agreement between the various 
 methods to estimate the background is reasonable.

\begin{table}[!h]
\begin{center}
\caption{\label{tab:final_iv} Backgrounds in the $m_{\mu \mu}$ and
  $m_X$ distributions from $e^+ e^- \rt \gamma \mu^+ \mu^-$ and $e^+
  e^- \rt \pp, \pp \rt (\gamma) \mu^+ \mu^-$ events that pass the
  selection criteria. Predicted results are from Monte Carlo
  simulations which do not include radiative return.  The $m_{\mu
    \mu}$ is determined above by fitting background in the $m_{\mu
    \mu}$ distribution.  Results are given with the radiative return
  determined from the difference between the background determined
  from the $\cos \theta_{J/\psi}$ distribution and the predicted $e^+
  e^- \rt \gamma \mu^+ \mu^-$ background
  added to the predicted and $m_{\mu \mu}$ results. These may be
  compared to the determinations obtained fitting the $\Delta \phi$
  distributions, which do contain radiative return.} \vspace{0.1in}
\begin{tabular} {|l|c|c|c|} \hline
Distribution & Predicted +
&  $m_{\mu \mu}$ + & $\Delta \phi$ \\
     & Meas. Rad. Ret.   & Meas. Rad. Ret.  & Dist. 
\\ \hline
$m_{\mu^+ \mu^-}$  & 1269 $\pm$ 122 & 1590 $\pm$ 140 & $1476 \pm 150$   \\ 
$m_X$               & 550 $\pm$ 80  & -- & $636 \pm 161$    \\ \hline
\end{tabular}
\end{center}
\end{table}

\subsection{\boldmath Exclusive Background}

The predicted background from $e^+ e^- \rt \gamma \mu^+ \mu^-$ and
$e^+ e^- \rt \pp$, $\pp \rt (\gamma) \mu^+ \mu^-$ in the $m_X$
distribution is smaller than that in the $\mu^+ \mu^-$
distribution.  However ``radiative return'' which is not
included in the Monte Carlo generator is not expected to be reduced by
the $\chi^2$ requirement that is made in going from the $m_{\mu^+ \mu^-}$
distribution to the $m_X$ distribution.

As was done above, the $\cos \theta_{J/\psi}$ distribution (but now
with a $\chi^2$ requirement) can be used to determine the amount of $e^+ e^-
\rt \gamma \mu^+ \mu^-$ background in the $m_X$ distribution.  Using a
similar calculation, the amount of background is determined to be 413
$\pm$ 75 events.
The measured background is larger
than the predicted background, and the
difference (221 $\pm$ 76) is taken as a measurement of the ``radiative
return''. Note that the two determinations of ``radiative return''
are consistent as expected.


The $\Delta \phi$ distributions with a $\chi^2$ requirement may be used, as
above, to determine the total background in the $m_X$ distribution
including ``radiative return''.  The result is $636 \pm 161$ events.
The amount of background when fitting the $m_X$ distributions will be
constrained to this amount.  Table~\ref{tab:final_iv} compares the
estimates and final backgrounds determined from $\Delta \phi$
distributions for both the $m_{\mu \mu}$ and $m_X$ backgrounds.  The
agreement between the various determinations for each distribution is
reasonable, and the total backgrounds are quite small.

\section{Fitting the mass distributions}

\subsection{Inclusive channel}

To determine the number of inclusive events, the $m_{\mu^+ \mu^-}$
distribution 
is fit with signal and background shapes.
The signal shape is obtained from real data using the 
$m_{\mu^+ \mu^-}$ distributions from
$\psi(2S) \rt {\rm anything} \: J/\psi$ (data) events with 
additional charged tracks.  These events are primarily 
due to $\pi^+ \pi^- J/\psi$.
For the background shape, the $m_{\mu^+ \mu^-}$ distribution
obtained by combining the distributions for Monte Carlo $e^+ e^- \rt
\gamma \mu^+ \mu^-$ and $e^+ e^- \rt \pp$, $\pp \rt (\gamma)\mu^+ \mu^-$
events is used.  Figure~\ref{fig:fit_mmumu} shows the fit to the
$m_{\mu^+ \mu^-}$ distribution.  The background distribution
differs somewhat from the data in the high mass region.  Because 
of this, the fit and background determinations everywhere in this
analysis are restricted to 
masses below 3.4 GeV/c$^2$.  The background in the 
$m_{\mu^+ \mu^-}$ distribution (Fig.~\ref{fig:fit_mmumu}) below 3.4 
GeV/c$^2$ is
determined to be 1307 $\pm$ 56 events.  The number of events in the
signal peak is 44498.

\subsection{Exclusive decays}

To determine the number of exclusive decays and separate $\psi(2S) \rt
\pi^0 \pi^0 J/\psi$ and $\psi(2S) \rt \pi^+ \pi^- J/\psi$ events,
$m_X$ histograms for events with and without additional charged
tracks, shown in  Figs.~\ref{fig:none}
and \ref{fig:one},
are fit simultaneously. 
There are 20818 and 19846 events in
the two distributions.  Contributions from the 
$\chi_{c0}$ are expected to be very small \cite{chic0} and are not 
included in the fit.  The influence of $\pp \rt \pi^0 J/\psi$ is also 
small, indeed there
is no indication of such a component in Fig.~\ref{fig:none}, and
this channel is also not included.  The $m_X$ distributions for 
$\chi_{c1}$, $\chi_{c2}$, and the combined background distribution 
are broad and rather similar in shape, as can be seen in 
Fig.~\ref{fig:none}.  Since these are difficult to distinguish,
the $\chi_{c2}$ to $\chi_{c1}$ ratio is constrained using calculated 
efficiencies and the PDG world average branching fractions for the two 
processes.  In addition, the amount of
background is constrained as discussed above.

In Fig.~\ref{fig:one}, it is seen that the $\chi_{c1}$, $\chi_{c2}$,
and $\eta$ contributions are small.  When fitting, the ratios of these
components to those in Fig.~\ref{fig:none} are also constrained using
Monte Carlo determined ratios. Since the amount of background is
predicted to be very small in this distribution, it is not included in
the fit. The numbers of fitted events obtained from the simultaneous
fits to the $m_X$ distributions of Figs.~\ref{fig:none} and
\ref{fig:one} and corresponding efficiencies are shown in
Table~\ref{tab:fitted}.

\begin{figure}[!htb]
\begin{center}
\begin{minipage}[t]{2.9in}
\centerline{\epsfysize 3.0 truein
\epsfbox{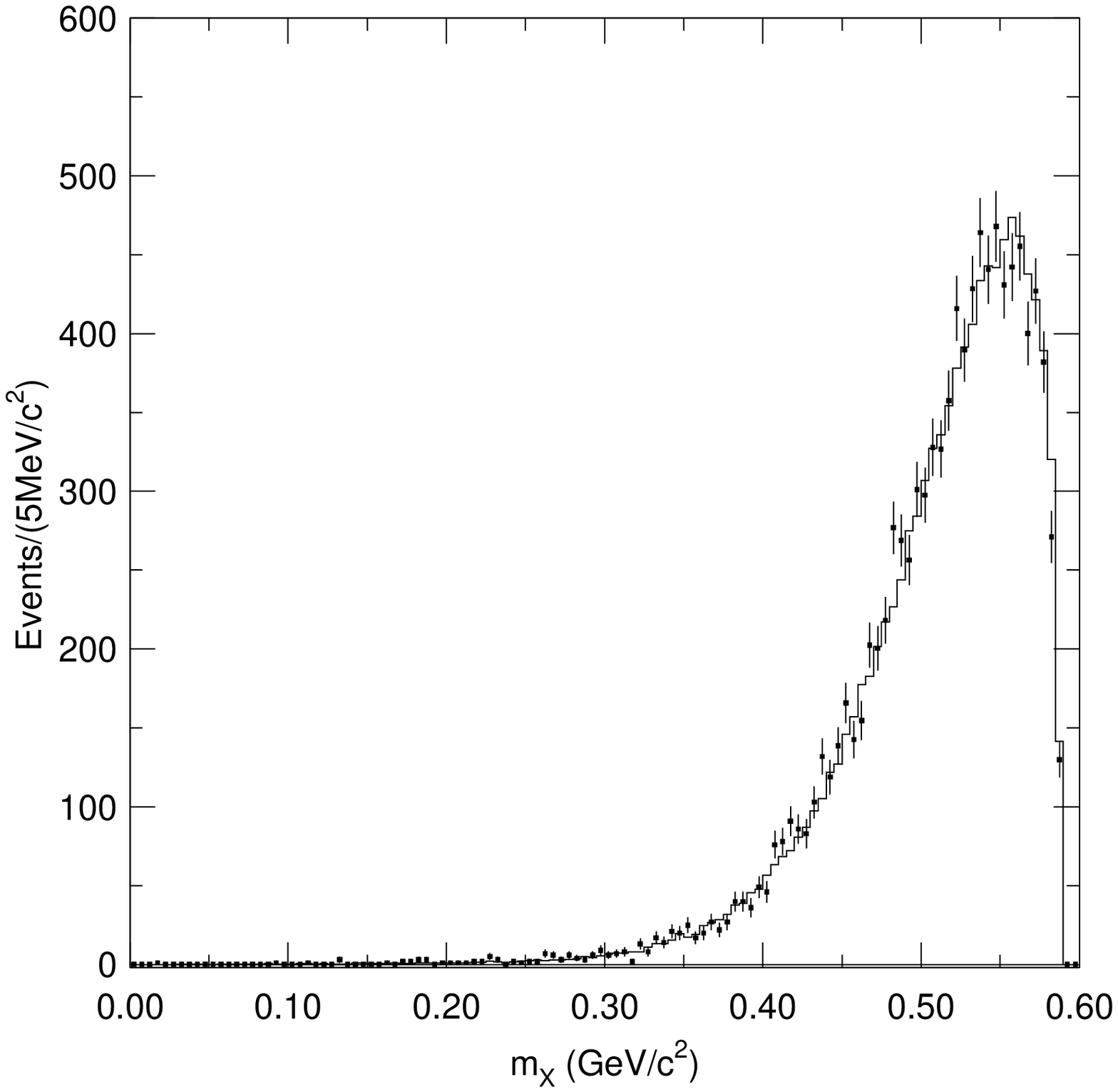}}
\caption{\label{fig:comp3} Data with two or more additional charged tracks
  with all Monte Carlo distributions, normalized by
  the fit result, subtracted except for
$\pi \pi$ (dots with error bars), fit with
$\pp \rt \pi \pi J/\psi$ Monte Carlo (solid histogram) plus a polynomial.}
\end{minipage} \ \
\begin{minipage}[t]{2.9in}
\centerline{\epsfysize 3.0 truein
\epsfbox{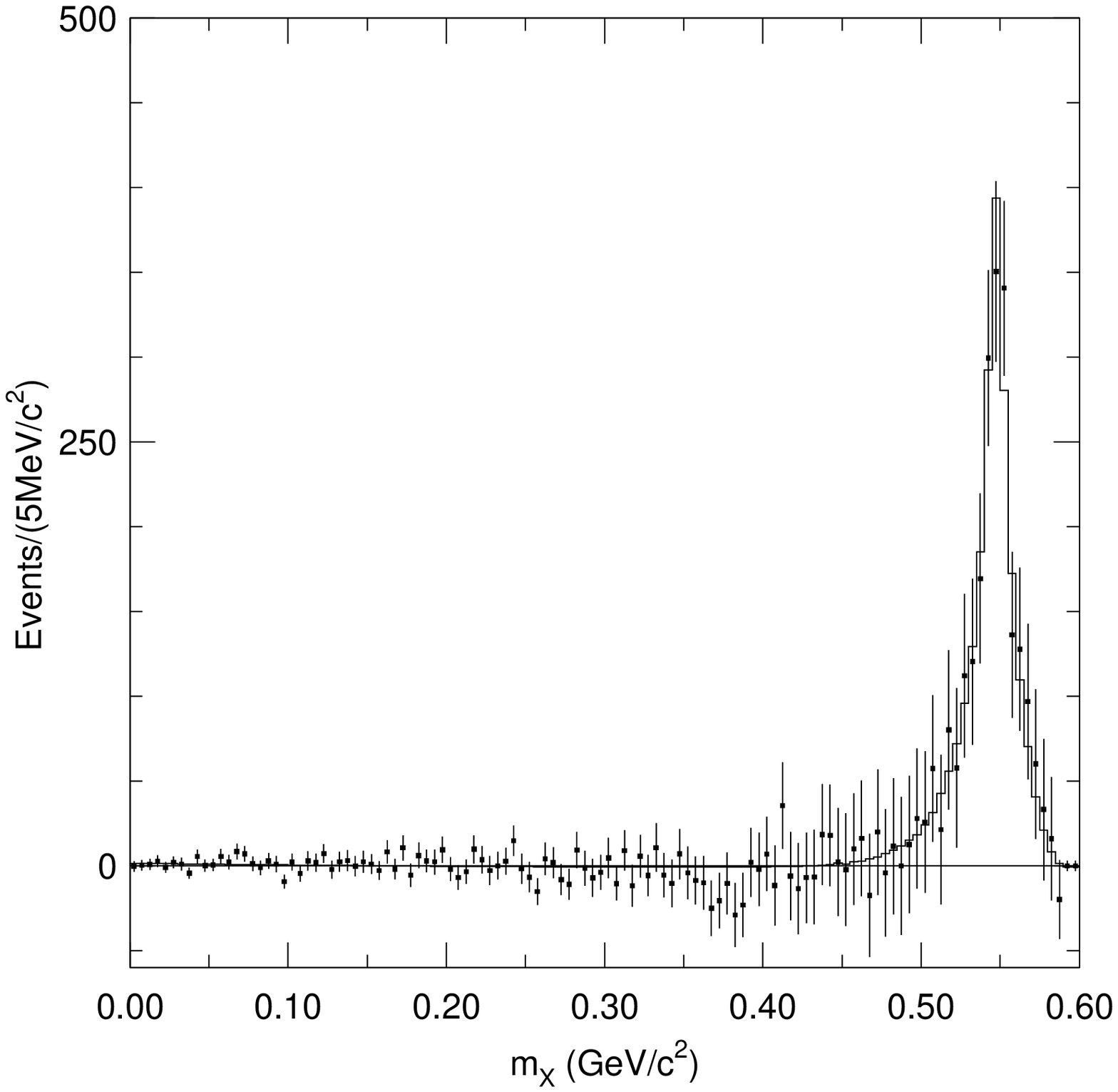}}
\caption{\label{fig:comp4} Data with no additional charged tracks with all
  Monte Carlo distributions, normalized by
  the fit result, subtracted except for
$\eta J/\psi$ (dots with error bars), fit with $\pp \rt \eta J/\psi$ Monte
Carlo (solid histogram) plus a polynomial.}
\end{minipage}
\end{center}
\end{figure}

As an indication as to how well the Monte Carlo distributions fit the
signal, Figs. \ref{fig:comp3} and \ref{fig:comp4} show the data
with all but one shape subtracted, normalized using the fit results,
and compared with the Monte Carlo distribution.

\begin{table}[!h]
\begin{center}
\caption{\label{tab:fitted}Fit results and efficiencies for inclusive
and exclusive decays.  The exclusive results are for fitting
the distributions shown in Figs.~\ref{fig:none}
and \ref{fig:one}
simultaneously. 
The numbers of events, $n$, and the efficiencies (eff)
for the exclusive cases are the results for the histogram in Fig.~\ref{fig:none}.
The zero error on the number of $\chi_{c2}$ events is
because of the $\chi_{c2}/\chi_{c1}$ constraint.}
\vspace{0.1in}
\begin{tabular} {|l|r|c|c|} \hline
 CASE   &        $n$  &  $\delta n$     &    eff  \\ \hline
Anything ($m_{\mu \mu}$)  &  44498          &      232      &    0.3488
 \\ \hline
 $\pi^+\pi^-$ and $\pi^0\pi^0$  &    13952  &   186 &   0.150  \\ 
 $\eta$    &     2121 &    117 &   0.3597  \\
 $\chi_{c1}$  &    2793 &   62  &   0.3678  \\
 $\chi_{c2}$  &    1326  &     0 &   0.3661  \\ \hline
\end{tabular}
\end{center}
\end{table}

\subsection{Determination of branching ratios}

Ratios of branching fractions, normalizing to the number of $\pp \rt
\pi^+ \pi^- J/\psi$ events, i.e.  $\frac{B(\psi(2S) \rt X
J/\psi)}{B(\psi(2S) \rt \pi^+ \pi^- J/\psi)}$ are calculated.  The
advantage of normalizing in this way is that many of the muon
selection systematic errors largely cancel, as well as the systematic error
due to the $\chi^2$ requirement.

First the correction factor 
for $\pi^0 \pi^0$ events in the $\pi \pi$ $m_X$ distribution with
additional charged tracks ($f_1$)
due to gamma conversions and
delta rays is determined:
$$f_{1} = \frac{0.305 \epsilon_{1}(\pi^+ \pi^-)}{0.305 \epsilon_{1}(\pi^+ \pi^-) +0.182 \epsilon_{1}(\pi^0 \pi^0)},$$
where $\epsilon_{1}(\pi^+ \pi^-)$ and $\epsilon_{1}(\pi^0 \pi^0)$
are the efficiencies for additional charged tracks for
$\pi^+ \pi^-$ and $\pi^0 \pi^0$ events, respectively, and 0.305 and
0.182 are the PDG branching fractions for $\pp \rt \pi^+ \pi^- J/\psi$
and $\pp \rt \pi^0 \pi^0 J/\psi$, respectively \cite{PDG02}. 
$$N_{1}(\pi^+ \pi^-) = f_{1} N_{1}(\pi \pi),$$
where  $N_{1}(\pi \pi)$ is the number of $\pi \pi$
events with additional charged tracks.  Since $\epsilon_{1}(\pi^0 \pi^0)$ is small,
$f_{1}$ is near 1.0: $f_{1} = 0.967$.

For $X = \eta$ or $X = \chi_c$,
$$\frac{B(\pp \rt X J/\psi)}{B(\pp \rt \pi^+ \pi^- J/\psi)}
=\frac{\epsilon_{{\rm ratio}1}\,\epsilon_{1}(\pi^+ \pi^-) N_0(X)}{
\epsilon_0(X) f_{1} N_{1}(\pi \pi)},$$ where $\epsilon_{{\rm ratio}1}$
is ratio of the track efficiencies for detecting one $\pi$ track for
data and Monte Carlo data, $N_0(X)$ is the number of
$X$ events in the histogram with no additional tracks, and $\epsilon_0(X)$
is the efficiency for $X$'s in the same sample.
\noindent
For the $\pi^0 \pi^0$ case:


$$\frac{B(\pp \rt \pi^0 \pi^0 J/\psi)}{B(\pp \rt \pi^+ \pi^- J/\psi)} =\frac{\epsilon_{{\rm ratio}1}\,\epsilon_{1}(\pi^+ \pi^-) N_0(\pi \pi)}{
\epsilon_0(\pi^0 \pi^0)   f_{1}   N_{1}(\pi \pi)} -
\frac{\epsilon_0(\pi^+ \pi^-)}{\epsilon_0(\pi^0 \pi^0)},$$
where $N_0(\pi \pi)$ is the number of events in the $\pi \pi$ peak in the
histogram with no additional tracks. 

For the $X \rt $ anything case,
the $m_{\mu \mu}$ distribution that is fitted to obtain the
number of $\pp \rt {\rm anything} \: J/\psi$ events, $N({\rm anything} \:
J/\psi)$, has no $\chi^2$ requirement.  Therefore a correction must be made for the
effect of this requirement since a $\chi^2$ requirement was made on the $\pp \rt \pi^+
\pi^- J/\psi$ distribution:

$$\frac{B(\pp \rt {\rm anything} \: J/\psi)}{B(\pp \rt \pi^+ \pi^- J/\psi)} =
\left(\frac{\epsilon_{{\rm ratio}1}\,\epsilon_{1}(\pi^+ \pi^-)}
{\epsilon({\rm anything} \: J/\psi)}\right)
\left( \frac{\epsilon_{\chi^2 {\rm cut}} N({\rm anything} \: J/\psi)|_{{\rm no~}
\chi^2 {\rm ~cut}}}{N_{1}(\pi^+ \pi^- J/\psi)}\right) $$

\noindent The efficiencies in the first fraction on the right of the
equal sign are determined from Monte Carlo and have a $\chi^2$ requirement.
The term $\epsilon({\rm anything} \: J/\psi)$ is the efficiency for the combination of
processes in $\pp \rt {\rm anything} \: J/\psi$.  The term
$\epsilon_{\chi^2 {\rm cut}}$ is the efficiency of the $\chi^2$ requirement
and is determined using data from the ratio of events with
additional tracks with and without
the $\chi^2$ requirement.  By requiring an extra charged track,
a clean sample of events with which to determine this ratio is selected.



The $m_X$ analysis is done using the distributions shown in Figs.~\ref{fig:none}
and \ref{fig:one}.  It may also be done using the distribution for
events with more than one additional charged tracks rather than the
one with any number of additional charged tracks.
Each has its advantages.  The first case has a higher
efficiency and should have a smaller percentage error on the
efficiency.  The second one should have less background from delta
rays and gamma conversions.
Our final results are determined from the averages 
of the
results for these two cases and
are summarized in Table~\ref{tab:results}.

\section{Systematic errors}

Systematic errors are summarized in Table~\ref{tab:sys}. Some of them
are determined by varying selection requirements. Other contributions
are determined by turning off the weighting used for $\pp \rt \pi \pi
J/\psi$ events, smearing the simulated $m_X$ by $\sigma =$ 3 MeV to test
the effect of changing the mass resolution, and removing
constraints. The amount of smearing of $m_X$ was determined by
comparing the width of a Gaussian fit to the $\eta$ peak for data in
Fig. \ref{fig:comp4} with a Gaussian fit to the corresponding Monte
Carlo distribution.  The data width is possibly 2 MeV wider (added in
quadrature) than the Monte Carlo width.

As shown in Fig.~\ref{fig:cospsi}b, the $\cos \theta_{J/\psi}$
distribution, where $\theta_{J/\psi}$ is the angle of the $J/\psi$ in
the lab, shows some peaking near $\pm 1.0$, which is indicative of
background from the radiative process.  To check whether this
background is being handled correctly, the effect of an additional requirement
on this angle ($|\cos \theta_{J/\psi}| < 0.9 $) is determined and
included in the systematic errors. The background estimate in the
$m_X$ distribution has also been increased by 30 \% to determine the
systematic error due to the uncertainty in the amount of background.

In the study done in Ref.~\cite{wenfeng},
the gamma conversion rate in data versus Monte Carlo data was
compared, and it was found that the rate was lower in data, (4.0 $\pm$
0.15) \%, than in Monte Carlo data, 4.5 \%.  The effect of this
difference has been studied using two different methods.  In the
first, the efficiencies for extra charged tracks in neutral decays is
changed by 4.0/4.5. In the second approach, a much higher track
momentum requirement ($p_{xy_{\pi}} > 0.15$ GeV/c) is made since gamma conversions
are predicted by the Monte Carlo simulation to be more important at
lower momentum.

This analysis has been done for two cases.  The first uses the
distribution with any number of additional charged tracks, while the second
uses the distribution with more than one additional charged tracks. For our
final results, the averages of the two sets of values are used, and
one half the difference between them is included in the systematic
errors.  Also included is an uncertainty of 2 \% for the uncertainty
in $\epsilon_{\rm ratio1}$.

\begin{table}[!h]
\begin{center}
\caption{\label{tab:sys}  Systematic error summary. The total error uses 1/2
the difference of the fitting results for the greater than zero and
the greater than one extra
charged track
cases in quadrature with all the other errors. Here $B(00)=B(\pp \rt
\pi^0 \pi^0 J/\psi)$, $B(+-) = B(\pp \rt \pi^+ \pi^- J/\psi)$,
  $B(\eta) = B(\pp \rt \eta J/\psi)$, $B(\chi_{cJ}) = B(\pp \rt \gamma
    \chi_{cJ})B(\chi_{cJ} \rt \gamma J/\psi)$, and $B({\rm anything})
    = B(\pp \rt {\rm anything} \: J/\psi)$. }
\vspace{0.1in}
\begin{tabular} {|lccccc|} \hline
Variation     & $\frac{B(00)}{B(+-)}$  &  $\frac{B(\eta)}{B(+-)}$ &
$\frac{B(\chi_{c1})}{B(+-)}$ & $\frac{B(
\chi_{c2})}{B(+-)}$ & $\frac{B({\rm anything})}{B(+-)}$ \\
               &   (\%)                &   (\%)   & (\%)& (\%)& (\%)\\  \hline
1.3 $\times ~m_X$ background         & 0.26    &  0.67    & 5.44   &
5.44 & -- \\ 
$p_{xy_{\pi}} > 0.08 \rt > 0.15$ GeV/c  & 0.88    &  0.34   &  0.74 &
0.51  & 0.22  \\
lower $\gamma$ conversion          & 1.11   &  1.01   & 0.74  &
1.04 & --\\
$|\cos \theta_{\pi}| < 0.8 \rt 0.75$ & 0.17  &  0.67   &  0.25 & 0.0 & 0.26\\
$|\chi_{dE/dx}|<3.0 \rt < 5.0$     & 1.93    & 0.67    &  0.99 &
1.04 & 0.73\\
$\chi^2 < 7 \rt < 10$               & 0.12   & 2.35    & 0.25  & 0.0
& 0.03\\
$|\cos \theta_{\mu}| < 0.6 \rt 0.65$ & 0.29  &  0.0  &  0.74 & 1.04
& 0.19\\
ymuid $> 1 \rt > 0 $               & 0.47    & 3.69    & 0.99  & 1.04
& 0.09\\
$|t_{\rm TOF}(\mu^+) - t_{\rm TOF}(\mu^-)| < 4 \rt < 5$
                    & 0.29    & 0.0    & 0.25  & 0.0 & 0.14\\
unweight $m_{\pi \pi}$             & 1.70    &  5.70    & 3.47  &
3.62 & 0.12 \\ 
smear mass (3 MeV)              & 0.70    & 4.36    & 0.74  & 0.51 &  0.42 \\
remove $\eta$ constraint           & 0.18    & 0.34    & 0.0  & 0.0 & 0.17\\
remove $\chi_{c2}$ constraint      & 1.70    & 2.35    &  29.0 &
46.1 & 0.45\\
$|\cos \theta_{J/\psi}| <0.9$      & 0.29    & 3.36    & 2.23  &
2.59 & 0.59 \\
$\epsilon_{\rm ratio1}$          & 2.0     &  2.0     &  2.0   &
2.0   &  2.0  \\
$> 0 \rt > 1$ extra tracks         &  4.17    & 3.05    & 3.17  &
 2.76 & 3.48\\ 
$m_{\mu^+ \mu^-}$ background $\rt$ $ e^+ e^- \rt \gamma \mu^+ \mu^-$ only  &   & & &  & 0.14 \\ \hline
Total                             & 4.58     & 9.82    & 29.9  & 46.8
& 2.92\\ \hline
\end{tabular}
\end{center}
\end{table}

Also indicated in Table~\ref{tab:sys} are the total systematic errors.
They are reasonably small except for $\pp \rt \gamma \chi_{c1}$ and
$\pp \rt \gamma \chi_{c2}$. The branching ratios are very different
when the  $\chi_{c2}$ constraint is removed. 

\section{Final results}

The final branching fraction ratios and branching fractions are shown
in Table~\ref{tab:results}, along with the PDG results, including
their experimental averages and global fit results.  For $B(\psi(2S)
\rt \pi^0 \pi^0 J/\psi)/B(\psi(2S) \rt \pi^+ \pi^- J/\psi)$, the PDG
does not use the previous experimental results and gives no average
value.  For the other four branching fraction ratios, only one
measurement exists for each, and Table~\ref{tab:results} lists the
single measurements quoted by the PDG. Our results for $B({\rm
anything} \: J/\psi)/B(\psi(2S) \rt \pi^+ \pi^- J/\psi)$ and $B(\eta
J/\psi)/B(\psi(2S) \rt \pi^+ \pi^- J/\psi)$ have smaller errors than
the previous results.  As a check, the sum of the exclusive branching
fraction ratios in the top of Table~\ref{tab:results} add to 1.854
$\pm$ 0.094, which is in good agreement with the inclusive branching
fraction ratio (1.867 $\pm$ 0.026), where the errors are the fit
errors only.

To determine the branching fractions, the ratios are multiplied by the PDG2002
value for $B(\pppp) = (30.5 \pm 1.6)$ \%.
The agreement for
both the ratios of branching fractions and the calculated branching
fractions using the PDG result for $\pp \rt B(\pi^+ \pi^- J/\psi)$
with the PDG fit results is
good. The $\chi_c$ results are high
compared to the PDG by about one sigma.

\begin{table}[!h]

\begin{center}
\caption{\label{tab:results} Final branching ratios and branching
  fractions.  PDG02-exp results are single measurements or
  averages of measurements, while PDG02-fit are results of their
  global fit to many expeimental measurements. For the value marked by
  $^*$, the PDG gives the reciprocal.  The BES results in the second
  half of the table are calculated using the PDG value of $B_{\pi \pi}
  = B(\pp \rt \pi^+ \pi^- J/\psi) = (30.5 \pm 1.6) \%$. } \vspace{0.1in}
\begin{tabular} {|l|c|c|c|} \hline
Case                                      &   This result &
PDG02-exp & PDG02-fit  \\ \hline
$B({\rm anything} \: J/\psi)/B_{\pi \pi}$   & $1.867  \pm 0.026  \pm 0.055$ & $2.016
\pm 0.150$ \cite{armstrong} &$1.828 \pm 0.036^*$   \\
$B(\pi^0 \pi^0 J/\psi)/B_{\pi \pi}$ & $0.570 \pm 0.009 \pm
0.026$ & - & $0.60 \pm 0.06$ \\
$B(\eta  J/\psi)/B_{\pi \pi}$ & $0.098 \pm  0.005 \pm
0.010$ & $0.091 \pm 0.021$ \cite{markii}& $0.103 \pm 0.010$\\
$B(\gamma \chi_{c1})B(\chi_{c1} \rt \gamma J/\psi)/B_{\pi \pi}$ & $0.126
\pm 0.003 \pm 0.038$ & $0.085 \pm 0.021$ \cite{markii} & $0.087 \pm 0.007$ \\ 
$B(\gamma \chi_{c2})B(\chi_{c2} \rt \gamma J/\psi)/B_{\pi \pi}$ & $0.060
\pm 0.000 \pm 0.028$ & $0.039 \pm 0.012$ \cite{markii} & $0.042 \pm 0.004$ \\ \hline
$B({\rm anything} \: J/\psi)$ (\%)  & $56.9  \pm 0.8 \pm 3.4  $ & $55 \pm 7$ & $55.7 \pm 2.6$ \\ 
 $B(\pi^0 \pi^0 J/\psi)$ (\%)  &  $17.4 \pm 0.3 \pm 1.2 $ & -- & $18.2 \pm 1.2$\\
 $B(\eta  J/\psi)$  (\%)    &  $3.00 \pm 0.16 \pm 0.33 $ & $2.9 \pm
0.5$ & $3.13 \pm 0.21$\\
 $B(\gamma \chi_{c1})B(\chi_{c1} \rt \gamma J/\psi)$ (\%)  & $3.9 \pm 0.1 \pm 1.2$
& $2.66 \pm 0.16$& $2.66 \pm 0.15$\\ 
 $B(\gamma \chi_{c2})B(\chi_{c2} \rt \gamma J/\psi)$ (\%)  & $1.84 \pm 0.01 \pm 0.86$
& $1.20 \pm 0.13$& $1.27 \pm 0.08$\\ \hline
\end{tabular}
\end{center}
\end{table}

\section{Summary}

In this analysis, the $m_{\mu^+ \mu^-}$ distribution for candidate
$J/\psi \rt \mu^+ \mu^-$ events is fit to determine $B(\psi(2S) \rt X
J/\psi, X \rt {\rm anything})$.  Energy - momentum conservation is
used to determine the recoil mass, $m_X$, and the $m_X$ distribution
is fit with Monte Carlo determined distributions to determine the
number of exclusive decays.  All processes are normalized to $\pppp$
to reduce systematic errors.
Ratios
of the branching fractions of $\pp \rt \eta J/\psi$, $\pi^0 \pi^0
J/\psi$, and anything $J/\psi$ to that of $\pp \rt \pi^+ \pi^-
J/\psi$ are measured to be $0.098 \pm 0.005 \pm 0.010$, $0.570 \pm
0.009 \pm 0.026$, and $ 1.867 \pm 0.026 \pm 0.055$, respectively.

\acknowledgments

   The BES collaboration thanks the staff of BEPC for their hard efforts.
This work is supported in part by the National Natural Science Foundation
of China under contracts Nos. 19991480,10225524,10225525, the Chinese Academy 
of Sciences under contract No. KJ 95T-03, the 100 Talents Program of CAS 
under Contract Nos. U-11, U-24, U-25, and the Knowledge Innovation Project of 
CAS under Contract Nos. U-602, U-34(IHEP); by the National Natural Science 
Foundation of China under Contract No.10175060(USTC),and No.10225522(Tsinghua University); 
and by the Department 
of Energy under Contract No.DE-FG03-94ER40833 (U Hawaii).


\noindent
\begin{thebibliography}{99}

\bibitem{PDG02} K. Hagiwara \etal, Phys. Rev. {\bf D66}, 010001 (2002).

\bibitem{suzuki} M. Suzuki, Phys. Rev. {\bf D63}, 054021 (2001).

\bibitem{gu} Y. F. Gu and X. H. Li, Phys. Rev. {\bf D63}, 114019 (2001).

\bibitem{ggjp} J. Z. Bai \etal, BES Collaboration, submitted to Phys. Rev. {\bf D}, hep-ex/0403023.

\bibitem{markI} G. S. Abrams \etal, Phys. Rev. Lett. {\bf 34}, 1181
  (1975); \\
J. S. Whitaker \etal, Phys. Rev. Lett. {\bf 37}, 1596
  (1976); \\
 W. M. Tanenbaum \etal, Phys. Rev. Lett. {\bf 36}, 402 (1976).

\bibitem{CNTR} C. J. Biddick \etal, Phys. Rev. Lett. {\bf 38}, 1324
  (1977); \\
W. Bartel \etal, Phys. Lett. {\bf B79}, 492 (1978).

\bibitem{DASP} R. Brandelik \etal, Nucl .Phys. {\bf B160}, 426
  (1979);\\
 R. Brandelik \etal, Z. Phys. {\bf C1}, 233 (1979).

\bibitem{markii} T. Himel \etal, Phys. Rev. Lett. {\bf 44}, 920 (1980).

\bibitem{crystalball} M. J. Oreglia \etal, Phys. Rev. Lett. {\bf 45},
  959 (1980); \\
M. J. Oreglia \etal, Phys. Rev. {\bf D25}, 2259 (1982); \\
J. Gaiser \etal, Phys. Rev. {\bf D34}, 711 (1986).


\bibitem{besI} J.Z. Bai et al., (BES Collab.), Nucl. Inst. Meth.
{\bf A344}, 319 (1994).


\bibitem{distributions} J. Z. Bai \etal, BES Collaboration, Phys. Rev. {\bf D62}, 32002 (2000).

\bibitem{mnfit} The fitting of the $m_{\mu \mu}$ and $m_X$
  distributions is done using Mn\_fit, www-zeus.physik.uni-bonn.de/$\sim$brock/mn\_fit.html.

\bibitem{mumu} For the $m_{\mu \mu}$ distribution, no $\chi^2$ requirement is
made but $m_{\mu \mu} < 3.4$ GeV/c$^2$ is used.

\bibitem{mx} Here $\chi^2 <7.0$ is also required.

\bibitem{chic0} From \cite{PDG02}, $B(\pp \rt \gamma \chi_{c0})
B(\chi_{c0} \rt \gamma J/\psi)$: $B(\pp \rt \gamma \chi_{c1})
B(\chi_{c1} \rt \gamma J/\psi)$: $B(\pp \rt \gamma \chi_{c2})
B(\chi_{c2} \rt \gamma J/\psi)$ = 0.026 : 1 : 0.477 





\bibitem{wenfeng}  J.Z. Bai \etal, BES Collaboration,
Phys. Rev. {\bf D67}, 052002 (2003). 

\bibitem{armstrong} T. A. Armstrong \etal, Phys. Rev. {\bf D55}, 1153 (1997). 








\end{thebibliography}
\end{document}